\documentclass[prd,preprintnumbers]{revtex4} 

\usepackage{graphicx}
\usepackage{dcolumn}
\usepackage{bm}
\usepackage{latexsym}
\usepackage{amsfonts}
\usepackage{amssymb}
\usepackage{amsmath}
\usepackage{bm}
\usepackage{color}

\begin{document}

\preprint{KUNS-2444}

\title{Anisotropic Non-Gaussianity from a Two-Form Field}

\author{Junko Ohashi$^1$}

\author{Jiro Soda$^2$}

\author{Shinji Tsujikawa$^1$}

\affiliation{$^1$Department of Physics, Faculty of Science, Tokyo University of Science, 1-3, Kagurazaka, Shinjuku, Tokyo 162-8601, Japan}

\affiliation{$^2$Department of Physics, Kyoto University, Kyoto 606-8502, Japan}

\date{\today}


\begin{abstract}

We study an inflationary scenario with a two-form field to which an inflaton 
couples non-trivially. First, we show that anisotropic inflation can be 
realized as an attractor solution and that the two-form hair remains 
during inflation. A statistical anisotropy can be developed because of
a cumulative anisotropic interaction induced by the background 
two-form field. The power spectrum of curvature perturbations
has a prolate-type anisotropy, in contrast to the vector models having an 
oblate-type anisotropy. We also evaluate the bispectrum and trispectrum
of curvature perturbations by employing the in-in formalism based on 
the interacting Hamiltonians.
We find that the non-linear estimators $f_{\rm NL}$ and $\tau_{\rm NL}$
are correlated with the amplitude $g_*$ of the statistical anisotropy 
in the power spectrum. Unlike the vector models, both $f_{\rm NL}$ and 
$\tau_{\rm NL}$ vanish in the squeezed limit.
However, the estimator $f_{\rm NL}$ can reach the order of 10 
in the equilateral and enfolded limits.
These results are consistent with the latest bounds
on $f_{\rm NL}$ constrained by Planck.

\end{abstract}

\maketitle


%
\section{Introduction}

The inflationary paradigm \cite{infpapers} can successfully account for 
the observed temperature fluctuations of the Cosmic Microwave 
Background (CMB) radiation and the distribution of large scale 
structures \cite{infper}.
The basic outcome of simplest single-field slow-roll inflation models is the statistical isotropy, the Gaussian and (almost) scale-invariant 
power spectrum \cite{review}. 
These statistical properties can be tested by precise measurements of 
the CMB temperature anisotropies.

The observational data provided by the Wilkinson Microwave Anisotropy Probe (WMAP) \cite{WMAP1} showed the evidence of 
the scale dependence of the power spectrum, whose property has been 
used to discriminate between a host of inflationary models.
Interestingly, the WMAP data also indicated the deviation from the 
Gaussian perturbations and they gave us a hint of the 
statistical anisotropy \cite{WMAP9}. 
According to the recent results of Planck, the Gaussian 
perturbations are still consistent with the data, 
but the statistical anisotropy remains \cite{Planckfirst}.
In the light of these new results, it is worth investigating a 
possible statistical anisotropy based on concrete theoretical models.  

Naively, the statistical anisotropy implies that vector fields can play an important role during inflation. 
A mechanism for creating the statistical anomaly at the end of inflation was proposed
in Ref.~\cite{Yokoyama} and extended in various ways~\cite{Dimopoulos}. 
Moreover,  a more concrete model has been proposed in the context of supergravity \cite{Watanabe}. 
There, the anisotropic inflation is realized as an attractor and 
the vector hair survives during inflation. 
The latter vector hair gives rise to rich phenomenology~\cite{Gumrukcuoglu:2010yc,Watanabe:2010fh,Emami:2013bk} such as the statistical anisotropy
 and the cross correlation between the curvature perturbations and the primordial gravitational waves 
(see Refs.~\cite{Soda:2012zm,Maleknejad} for reviews).  
In particular, the latter would imply the correlation 
between the temperature fluctuation and the 
{\rm B}-mode polarization \cite{Watanabe2}.

Remarkably, subsequent works revealed that vector fields can also induce the large non-Gaussianity~\cite{BarnabyPRL}. 
In particular, the non-Gaussianity of curvature perturbations has been investigated in the context of anisotropic inflation ~\cite{Barnaby}-\cite{Baghram:2013lxa}, which further emphasized the rich phenomenology of the anisotropic inflationary models~\cite{Kanno:2010nr}. 
As a result, the Planck constrained the anisotropic inflationary models with vector fields strongly.  
In Ref.~\cite{Watanabe:2010fh}, however, it was pointed out that not only vector fields but also
two-form fields can potentially give rise to anisotropic inflation.
In fact, it is well known that there are two-form fields in string theory \cite{LCW}.
Hence it is natural to explore this possibility from the theoretical 
point of view. One may suspect that there is no statistical anisotropy because 
the two-form field can be represented by a pseudo scalar field, i.e., axion. However, there remains a possibility that a non-trivial polarization of 
a two-form field induces the statistical anisotropy. 
In this case the anisotropy comes from an expectation value of the spatial derivative of the axion field. At first glance this seems to be odd, 
but it is a natural framework from the picture of two-form fields. 
Hence, in this paper, we study this possibility in detail. 

We show that several interesting features are present in our model. 
Analogous to the results of Ref.~\cite{Watanabe}, 
anisotropic inflation can be sustained by the background two-form field.
Moreover, as suggested in Ref.~\cite{Watanabe:2010fh}, 
there exists a prolate type anisotropy 
in the power spectrum of curvature perturbations.
In contrast to vector models the non-Gaussianity vanishes 
in the squeezed limit, but the nonlinear estimator $f_{\rm NL}$
of the equilateral and enfolded shapes can be as large as the order of 10.
Hence, our predictions are consistent with the Planck data and the future analysis may reveal these statistical anisotropies. 

The organization of our paper is as follows. 
In Sec.~\ref{backsec}, we introduce the two-form field model 
and explain how the non-trivial hair remains during inflation. 
In Sec.~\ref{persec}, we quantize the perturbations of
two-form fields and derive their vacuum expectation values on 
super-Hubble scales. 
In Sec.~\ref{nonsec}, we evaluate various statistical 
quantities--such as $n$-point correlation functions ($n=2,3,4$) 
of curvature perturbations.  
Sec.~\ref{consec} is devoted to conclusions.


%
\section{Anisotropic Background}
\label{backsec}

We study the background dynamics of anisotropic inflation in the presence of 
two-form fields. The analysis is similar to the case of vector 
models studied in Ref.~\cite{Watanabe}, but we repeat it for completeness. 
Here, we emphasize the shape of anisotropy is different from that 
of vector models.  

Let us start with the action
\begin{equation}
S = \int d^4 x \sqrt{-g} \left[\frac{M_p^2}{2}R 
- \frac{1}{2}\partial^\mu \phi \partial_\mu \phi - V(\phi )
-\frac{1}{12}f^2 (\phi ) H^{\mu\nu\lambda} H_{\mu\nu\lambda} \right]\,,
\label{oriaction}
\end{equation}
where $M_p$ is the reduced Planck mass, $R$ is a scalar curvature 
calculated from a metric $g_{\mu\nu}$ (with a determinant $g$), 
$\phi$ is an inflaton field with a potential $V(\phi)$, 
and the field $H_{\mu\nu\lambda}$ has a non-trivial 
coupling $f(\phi)$ with the inflaton. 
The field $H_{\mu\nu\lambda}$ is related to 
a two-form field $B_{\mu \nu}$, as 
\begin{equation}
H_{\mu\nu \lambda} = 
\partial_\mu B_{\nu \lambda} + \partial_\nu B_{\lambda\mu}
+ \partial_\lambda B_{\mu\nu}\,.
\label{Hdef}
\end{equation}
Without loss of generality, one can take the $(y,z)$ plane 
in the direction of the two-form field. 
Then we can express $B_{\mu \nu}$ in the form
\begin{equation}
\frac{1}{2}B_{\mu\nu}\,dx^{\mu}\wedge dx^{\nu}= 
v(t)\,dy \wedge dz\,,  
\end{equation}
where $v(t)$ is a function with respect to the 
cosmic time $t$.

Since there exists a rotational symmetry in the $(y,z)$ plane, 
it is convenient to parameterize the metric as follows:
\begin{equation}
ds^2 = -{\cal N}(t)^2dt^2 + e^{2\alpha(t)} \left[ e^{-4\sigma (t)}dx^2
+e^{2\sigma (t)}(dy^2 + dz^2) \right] \ ,
\end{equation}
where $\alpha$ describes the average expansion 
(the number of e-foldings) and $\sigma$
represents the anisotropy of the Universe.
Here, the lapse function ${\cal N}$ is introduced 
to derive the Hamiltonian constraint.
With the above ansatz, the action (\ref{oriaction}) reduces to
\begin{equation}
S=\int d^4x \frac{1}{\cal N}e^{3\alpha} \left[ 
3 M_p^2(-\dot{\alpha}^2+\dot{\sigma}^2)
+\frac{1}{2}\dot{\phi}^{2}-{\cal N}^2 V(\phi)
+\frac{1}{2}f(\phi )^2\dot{v}^2 e^{-4\alpha (t) -4\sigma(t) } \right],
\end{equation}
where an overdot denotes a derivative with respect to $t$.
After taking variations, we set the gauge ${\cal N}=1$.

The equation of motion for the two-form field $v$ is easily solved as
\begin{equation}
\dot{v} =  A f(\phi)^{-2}  e^{\alpha +4\sigma}\,,
\label{eq:Ax}
\end{equation}
where $A$ is a constant of integration. 
Taking the variation of the action with respect to
${\cal N}, \alpha, \sigma, \phi$ and using the solution (\ref{eq:Ax}), 
we obtain the following background equations of motion
\begin{eqnarray}
\dot{\alpha}^2 &=& \dot{\sigma}^2+\frac{1}{3M_p^2} 
\left[ \frac{1}{2}\dot{\phi}^2+V(\phi )
+\frac{A^2}{2}f(\phi )^{-2}e^{-2\alpha + 4\sigma} \right]\,, 
\label{eq:hamiltonian}\\
\ddot{\alpha} &=& -3\dot{\alpha}^2+ \frac{1}{M_p^2} \left[ V(\phi) 
+\frac{ A^2}{3}f(\phi )^{-2}e^{-2\alpha+4\sigma} \right] \,,
\label{eq:alpha}\\
\ddot{\sigma} &=&  -3\dot{\alpha}\dot{\sigma} 
- \frac{A^2}{3M_p^2}f(\phi)^{-2}e^{-2\alpha +4\sigma} \ , 
\label{eq:sigma}\\
\ddot{\phi} &=& -3\dot{\alpha}\dot{\phi}-V_{,\phi}
+A^2f(\phi)^{-3} f_{,\phi}\,e^{-2\alpha + 4\sigma} 
\label{eq:inflaton} \ ,
\end{eqnarray}
where the subscript in $V_{,\phi}$ and $f_{,\phi}$ denotes a derivative 
with respect to its argument $\phi$.

Let us introduce the energy density of the two-form field
\begin{equation}
\label{rho-v-section-4}
\rho _v \equiv \frac{A^2}{2}f(\phi)^{-2}e^{-2\alpha +4\sigma}\,,
\end{equation}
and the shear $\Sigma \equiv \dot{\sigma}$. 
We also define the expansion rate $H\equiv \dot{\alpha}$.
First, we need to look at the shear to the expansion rate ratio $\Sigma/H$, 
which characterizes the anisotropy of the inflationary Universe.
Notice that Eq.~(\ref{eq:sigma}) reads
\begin{equation}
\dot{\Sigma} = -3H\Sigma 
- \frac{2}{3}\frac{\rho_v}{M_p^2} \ .
\end{equation}
In the regime where $\dot{\Sigma}$ becomes negligible, 
the ratio $\Sigma/H$ should converge to the value
\begin{equation}\label{R-def}
\frac{\Sigma}{H} = - \frac{2}{3}\frac{\rho _v}
{V(\phi)} \ ,
\end{equation}
where we used Eq.~(\ref{eq:hamiltonian}) 
under the slow-roll approximation, i.e.,
\begin{equation}
\dot\alpha^2=H^2 \simeq \frac{V(\phi)}{3M_p^2}\,.
\label{slow-roll:hamiltonian}
\end{equation}

In order for the anisotropy to survive during inflation,
$\rho_v$ must be almost constant.
Employing the standard slow-roll approximation and assuming
that the two-form field is sub-dominant in the inflaton equation 
of motion (\ref{eq:inflaton}), 
one can show the coupling function $f(\phi)$ should be
proportional to $e^{-\alpha}$ to keep $\rho_v$ almost constant.
In the slow-roll regime, the number of e-foldings $\alpha$ 
is related to $\phi$, as $d\alpha = -V(\phi)\,d\phi/(M_p^2 V_{,\phi})$.
Then the functional form of $f(\phi)$ is determined as
\begin{equation}
f(\phi) = e^{-\alpha} = e^{ \int \frac{V}{M_p^2 V_{,\phi}} d\phi} \ .
\label{eq:function}
\end{equation}
For the polynomial potential $V\propto \phi ^n$, for example, 
we have $f=e^{\frac{ \phi ^2}{2nM_p^2}}$. 
The above case is, in a sense, a critical one. 
What we want to consider is super-critical cases where 
the energy density of the two-form field increases.
For simplicity, we parameterize $f(\phi)$ by
\begin{equation}
f(\phi) = e^{c \int \frac{V}{M_p^2 V_{,\phi}}d\phi}\,,
\label{formula:f}
\end{equation}
where $c$ is a constant parameter. 
Let us consider the super-critical cases $c>1$.
From the definition (\ref{formula:f}), 
we can derive the following relation
\begin{equation}
\frac{f_{,\phi}}{f} = c  \frac{V}{M_{p}^2V_{,\phi}}
\label{formula:f1}\ .
\end{equation}
Then, the condition $c>1$ translates into
\begin{equation}
\frac{f_{,\phi}}{f} \frac{M_p^2 V_{,\phi}}{V} >1  \ .
\label{general-condition}
\end{equation}
Thus, any functional pairs of $f$ and $V$ satisfying the condition 
(\ref{general-condition}) in some range could produce 
the two-form hair during inflation.

On using Eq.~(\ref{formula:f1}), the inflation equation (\ref{eq:inflaton}) reads
\begin{equation}
\ddot{\phi} = -3\dot{\alpha}\dot{\phi}-V_{,\phi} \left[ 1-\frac{c}{\epsilon_V}
\frac{\rho_v}{V(\phi)}\right] \ , \label{eq:inflaton2}
\end{equation}
where the slow-roll parameter $\epsilon_V$ is defined as 
\begin{equation}
\epsilon_{V} \equiv \frac{M_p^2}{2} 
\left( \frac{V_{,\phi}}{V} \right) ^2 \ .
\label{epsilon-eta-def}
\end{equation}
In this case, if the two-form field is initially small 
(${\rho_v}/{V(\phi)} \ll \epsilon_V /c$), then the conventional 
single-field slow-roll inflation is realized. 
During this stage $f\propto e^{-c\alpha}$ and the energy density of the two-form field grows 
as $\rho_v \propto e^{2(c-1)\alpha}$. 
Therefore, the two-form field eventually becomes relevant to 
the inflaton dynamics described by Eq.~(\ref{eq:inflaton2}). 
Nevertheless, the cosmic acceleration continues because  ${\rho_v}/{V(\phi)}$ does not exceed $\epsilon_{V}/c$. In fact, if ${\rho_v}/{V(\phi)}$ exceeds $\epsilon_{V}/c$, the inflaton field $\phi$
does not roll down, which makes $\rho _v = A^2 f(\phi)^{-2}e^{-2\alpha+4\sigma} /2$
decrease. Hence the condition $\rho_v \ll V(\phi)$ is always satisfied.
In this way, there appears an attractor where inflation continues even 
when the two-form field affects the inflaton dynamics \cite{Watanabe}.

Let us make the above statement more precise.
Under the slow-roll approximation, the inflaton equation of motion 
(\ref{eq:inflaton}) reads
\begin{equation}
-3\dot{\alpha}\dot{\phi}-V_{,\phi}+A^2f^{-3}f_{,\phi}
e^{-2\alpha +4\sigma}  \simeq 0 \ .
\label{eq:inflaton3}
\end{equation}
Using Eqs.~(\ref{slow-roll:hamiltonian}) and (\ref{eq:inflaton3}), 
we obtain 
\begin{equation}
\frac{d \phi}{d \alpha}=\frac{\dot{\phi}}{\dot{\alpha}}
=-\frac{M_p^2 V_{,\phi}}{ V} + c \frac{A^2}{V_{,\phi}}
e^{-2\alpha +4\sigma -2c  \int\frac{V}{M_p^2 V_{,\phi}}d\phi} \ .
\label{eq:inflaton4}
\end{equation}
Neglecting the evolution of $V$, $V_{,\phi}$ and $\sigma$, 
this equation can be integrated to give
\begin{eqnarray}
e^{2\alpha -4\sigma +2c  \int\frac{V}{M_p^2 V_{,\phi}}d\phi}
= \frac{c^2 A^2}{c-1} \frac{V}{M_p^2 V_{,\phi}^2}
\left[ 1+ \Omega\,e^{-2(c-1)\alpha -4\sigma} \right]  \ ,
\label{exponential}
\end{eqnarray}
where $\Omega$ is a constant of integration.
Substituting this back into the slow-roll equation 
(\ref{eq:inflaton4}), it follows that 
\begin{equation}
\frac{d\phi}{d\alpha} = - \frac{M_p^2 V_{,\phi}}{V}+\frac{c-1}{c}
\frac{M_p^2 V_{,\phi}}{V} \left[ 1+ \Omega\,e^{-2(c-1)\alpha-4\sigma}
\right]^{-1} \,.
\label{dphi}
\end{equation}
At the initial stage of inflation ($\alpha \rightarrow -\infty$), 
the second term of Eq.~(\ref{dphi}) can be neglected relative 
to the first term.
In the asymptotic future ($\alpha \rightarrow \infty$),
the term containing $\Omega$ disappears.
This clearly shows a transition from the conventional 
slow-roll inflationary phase, where
\begin{equation}
\frac{d\phi }{d\alpha} = - \frac{M_p^2V_{,\phi}}{ V}\,,
\label{first-stage}
\end{equation}
to what we refer to as the second inflationary phase, where the two-form field is relevant to the inflaton dynamics and the inflaton gets $1/c$ times slower as
\begin{equation}
\frac{d\phi }{ d\alpha} = - \frac{1}{c}
\frac{M_p^2 V_{,\phi}}{V}\,.
\label{second-stage}
\end{equation}

In the second inflationary phase, we can employ Eq.~(\ref{exponential}) with discarding the $\Omega$ term to rewrite the energy density 
of the two-form field as
\begin{equation}
\rho _v =\frac{A^2}{2} e^{-2\alpha+4\sigma-2c \int\frac{V}
{M_p^2 V_{,\phi}}d\phi}
 =  \frac{c-1}{c^2} \epsilon_{V} V(\phi)\,,
\end{equation}
which yields the anisotropy
\begin{equation}
\frac{\Sigma}{H}= - \frac{2}{3} \frac{c-1}{c^2} \epsilon_{V} \ .
\end{equation}
Moreover, from Eqs.~(\ref{eq:hamiltonian}) and (\ref{eq:alpha}), 
the slow-roll parameter $\epsilon \equiv -\dot{H}/H^2$ is related 
to $\epsilon_V$ as
\begin{equation}
\epsilon =-\frac{\ddot{\alpha}}{\dot{\alpha}^2}=-\frac{1}{\dot{\alpha}^2} \left(-\frac{1}{2}  \dot{\phi}^2-\frac{1}{3} \rho_v \right) = \frac{1}{c} \epsilon_{V} \ ,
\end{equation}
where we neglected the anisotropy and used 
Eqs.~(\ref{slow-roll:hamiltonian}) and (\ref{second-stage}).
Thus we arrive at the result
\begin{equation}\label{Sigma-over-H-generic}
\frac{\Sigma}{H} = - \frac{2}{3}\frac{c-1}{c}\epsilon.
\end{equation}
Therefore, for a broad class of inflaton potentials and 
two-form kinetic functions, there exist anisotropic inflationary 
solutions, with $\Sigma/H$ proportional to $\epsilon$.
We also confirmed the existence of such anisotropic 
solutions by integrating 
Eqs.~(\ref{eq:hamiltonian})-(\ref{eq:inflaton}) numerically.
Note that the sign of $\Sigma/H$ is opposite to that 
derived for the vector field \cite{Watanabe}.

For $c=1$, we need a separate treatment. 
In this case, integration of Eq.~(\ref{eq:inflaton4}) gives
\begin{eqnarray}
e^{2\alpha -4\sigma +2 \int\frac{V}{M_p^2 V_{,\phi}}d\phi}
=2 A^2 \frac{V}{M_p^2 V_{,\phi}^2} (\alpha + \alpha_0 )\,,
\end{eqnarray} 
where $\alpha_0$ is an integration constant. 
Thus, we obtain the anisotropy
\begin{equation}
\label{Sigma-over-H-generic2}
\frac{\Sigma}{H} = 
- \frac{1}{3(\alpha +\alpha_0 )}\epsilon \ .
\end{equation}

Notice that the anisotropy stemmed from the two-form field is a prolate 
type, while the anisotropy created by the vector field was an oblate type. 
This can be understood by the fact that the vector extending to the $x$-direction speeds down the expansion in that direction, 
while the two-form field extending in the $(y,z)$ plane 
speeds down the expansion in the $(y,z)$ direction.  

Before entering the study of perturbations, we should note the following 
important attractor mechanism~\cite{Kanno:2009ei}. 
Taking a look at the result (\ref{second-stage}), 
we find that the relation
\begin{equation}
f = e^{c  \int\frac{V}{M_p^2 V_{,\phi}}d\phi} \propto e^{-\alpha}\,,
\end{equation}
holds during the second anisotropic inflationary phase. 
Recall that this is the critical behavior. 
As we will see in the next section, this attractor gives rise to 
the scale-invariant spectrum of the two-form field.


%
\section{Perturbations of two-form fields}
\label{persec}

From the phenomenological point of view the anisotropy of the expansion rate
needs to be sufficiently small, so it is a good approximation to neglect 
the effect of the anisotropic expansion.
However, we cannot ignore the effect of the two-form hair.
Actually, in the next section, we will show several interesting results. 
In this section, we prepare some tools for the evaluation of 
$n$-point correlation functions of curvature perturbations.

Let us consider the two-form field given by the action
\begin{equation}
S_{\rm int} = -\frac{1}{12}\int d^4 x \sqrt{-g}\, f^2 (\phi ) 
H^{\mu\nu\lambda} H_{\mu\nu\lambda} \,.
\label{actionani}
\end{equation}
In the above action, there exists a gauge invariance
under the gauge transformation
\begin{equation}
\delta B_{\mu\nu} = \partial_\mu 
\xi_\nu - \partial_\nu \xi_\mu \ .
\end{equation}
Here, since we have the redundancy 
$\xi_\mu \rightarrow \xi_\mu + \partial_\mu {\cal X}$, 
the parameter $\xi$ can be restricted to be $\partial_i \xi_i=0$.

For the derivation of the perturbation equations of the two-form 
field, we consider the isotropic background described by the metric
\begin{equation}
ds^2 = a^2 (\tau) \left(
-d\tau^2 + \delta_{ij} dx^i dx^j \right)\,,
\end{equation}
where $\tau=\int a^{-1} dt$ is the conformal time.
{}From Eq.~(\ref{Hdef}) it is convenient to perform the (3+1)-decomposition
\begin{eqnarray}
H_{0ij} &=& B'_{ij} + \partial_i B_{j0} + \partial_j B_{0i} \ , \\
H_{ijk} &=&  \partial_i B_{jk} + \partial_j B_{ki} + \partial_k B_{ij}\,,
\end{eqnarray}
where a prime represents a derivative with respect to $\tau$.
The interacting action (\ref{actionani}) reads
\begin{eqnarray}
S_{\rm int}=-\frac{1}{4}\int d^4 x \sqrt{-g}\, 
f^2 (\phi) H^{0ij} H_{0ij} 
-\frac{1}{12}\int d^4 x \sqrt{-g}\, f^2 (\phi ) H^{ijk} H_{ijk}\,. 
\end{eqnarray}
Taking the variation of this action with respect to $B_{0i}$, 
we obtain
\begin{equation}
\partial_i \left[ \sqrt{-g}\, f^2 (\phi ) H^{0ij} \right] = 0\,.
\label{master1}
\end{equation}
Variation with respect to $B_{ij}$ leads to
\begin{equation}
\frac{\partial}{\partial \tau} 
\left[ \sqrt{-g}\, f^2 (\phi) H^{0ij} \right]
+ \partial_k \left[ \sqrt{-g}\, f^2 (\phi) H^{ijk}  \right]  
= 0\,.
\label{master2}
\end{equation}
Then Eqs.~(\ref{master1}) and (\ref{master2}) reduce to
\begin{eqnarray}
&&  \partial_i \left[ B'_{ij} + \partial_i B_{j0} 
- \partial_j B_{i0} \right]  = 0\,, 
\label{Beq-1}\\
&& - \frac{\partial}{\partial \tau} \left[ \frac{1}{a^2}f^2 (\phi) 
\left\{ B'_{ij} + \partial_i B_{j0} + \partial_j B_{0i} \right\} \right]
+ \partial_k \left[ \frac{1}{a^2} f^2 (\phi ) 
\left\{ \partial_i B_{jk} + \partial_j B_{ki} 
+ \partial_k B_{ij} \right\}  \right]=0\,.
\label{Beq-2}
\end{eqnarray}

Since we have the gauge degree of freedom
\begin{equation}
\delta B_{ij} = \partial_i \xi_j - \partial_j \xi_i  \ , 
\qquad \partial^i \xi_i =0 \ ,
\end{equation}
choosing the parameter
\begin{equation}
\Delta\, \xi_j = - \partial_i B_{ij}  \ ,
\end{equation}
we can take the gauge
\begin{equation}
\partial_i B_{ij} =0 \,.
\end{equation}
Similarly, by taking into account the following gauge 
transformation
\begin{equation}
\delta B_{i0}=\partial_i \xi_0 
- \xi'_i\,,
\end{equation}
we can set
\begin{equation}
\partial_i B_{i0}=0\,.
\end{equation}
{}From Eq.~(\ref{Beq-1}) it follows that 
\begin{equation}
B_{i0} =0 \,.
\end{equation}
Then, Eq.~(\ref{Beq-2}) reduces to 
\begin{equation}
-\frac{\partial}{\partial \tau} \left[ \frac{1}{a^2} 
f^2 (\phi) B'_{ij} \right]
+ \frac{1}{a^2} f^2 (\phi) \partial_k^2 B_{ij} = 0\,.
\label{pB-eq}
\end{equation}

Let us check the degrees of freedom. The components $B_{ij}$ have 3 degrees.
There are 2 gauge conditions $B_{ij,j}=0$. Hence, we have one degree of freedom.
This is the reason why we can map the two-form field to a scalar field.
However, there is an important difference from the scalar field, 
namely, there exists a polarization in our case.
Using the polarization tensor $\epsilon_{ij} = - \epsilon_{ji}$ 
with $k_i \epsilon_{ij} =0$, we can expand the anti-symmetric 
tensor field as
\begin{equation}
B_{ij}(\tau, {\bm x})= \int \frac{d^3 k}{(2\pi)^{3/2}} 
\left[ a_{\bm k}\, \chi (\tau, {\bm k})\, \epsilon_{ij}\, 
e^{i{\bm k} \cdot {\bm x}} + {\rm c.c.} 
\right] \,,
\end{equation}
where $a_{\bm k}$ is the annihilation operator.
We take the polarization tensor to be 
$\epsilon_{ij} = (k_m/k)\epsilon_{mij}$, with 
the normalization condition
\begin{equation}
\epsilon_{ij} \epsilon^{ij} = 2\,.
\end{equation}

The mode functions $\chi (\tau, {\bm k})$ is known by solving 
Eq.~(\ref{pB-eq}). 
For convenience we introduce the following variable
\begin{eqnarray}
u (\tau, {\bm k})=\frac{f}{a}\, \chi (\tau, {\bm k})\,,
\end{eqnarray}
and parametrize the kinetic function as $f=a^p $.
This is always possible during inflation where both 
$\phi$ and $a=e^\alpha$ are monotonic functions
with respect to time.
Then, we obtain
\begin{eqnarray}
u''+\left[ k^2 + \frac{p (1-p )}{\tau^2} \right] u = 0\,.
\end{eqnarray}
Hence, for $p = -1$ or $p = 2$, we obtain the scale-invariant spectrum
for the two-form field.
Although either choice is possible, we set $p =-1$ because, 
as is shown in the previous section, 
this is an attractor value realized during anisotropic inflation. 
For $p=-1$, we can deduce the mode functions as
\begin{eqnarray}
u=  \frac{H a}{\sqrt{2k^3}} 
\left( 1+ik\tau \right) e^{-ik\tau} \,, \label{uso}
\end{eqnarray}
and $\chi = a^2 u$. 

Now, it is convenient to define 
\begin{eqnarray}
E_{yz} &\equiv& \frac{f}{a^3} H_{0yz}
=\frac{f}{a^3}B_{yz}'\,,\label{Edef1}\\
\delta E_{ij} &\equiv& \frac{f}{a^3} H_{0ij}
= \frac{f}{a^3}B_{ij}' \label{Edef2}\,,  
\end{eqnarray}
where $E_{yz}$ and $\delta E_{ij}$ correspond to the background value and 
the perturbation of the two-form field respectively.
Note that $H_{ijk}$ can be negligible on super-Hubble scales.
We perform the Fourier transformation of the perturbation 
$\delta E_{ij}$, as
\begin{equation}
\delta E_{ij}=\int \frac{d^3 k}{(2\pi)^{3/2}}
e^{i {\bm k}\cdot {\bm x}}\,\delta {\cal E}_{ij} (\tau,{\bm k})\,.
\label{delEdef}
\end{equation}
On super-Hubble scales the Fourier modes are given by 
\begin{eqnarray}
\delta {\cal E}_{ij}(\tau ,{\bm k}) = \left( a_{\bm k} 
+ a^{\dagger}_{-{\bm k}}  \right) 
{\cal E}_k\,\epsilon_{ij} \ ,
\end{eqnarray}
where 
\begin{eqnarray}
{\cal E}_k =\frac{f}{a^3}\chi'=\frac{(a^2 u)'}{a^4}
\simeq \frac{3H^2}{\sqrt{2k^3}}\,.
\label{calE}
\end{eqnarray}
In the last approximate equality of Eq.~(\ref{calE}) 
we used the solution (\ref{uso})
in the limit $\tau \to 0$ on the de Sitter background.

The vacuum expectation value of the field $\delta E_{ij}$
is given by 
\begin{equation}
\langle  \delta E_{ij}^2 \rangle = \frac{1}{\pi^2}
\int dk~k^2 |{\cal E}_k|^2 \simeq 
\frac{9H^4}{2\pi^2} \int_{\rm IR} \frac{dk}{k}\,.
\end{equation}
The Infrared (IR) modes are characterized by 
$k_i<k<k_f$, where $k_i$ and $k_f$ are the wavenumbers
which crossed the Hubble radius at the beginning and 
at the end of inflation respectively.
Since the integral $\int_{\rm IR} \frac{dk}{k}=\ln (k_f/k_i)$ 
is equivalent to the number of e-foldings $N=\ln (a_f/a_i)$
on the de-Sitter background, it follows that 
\begin{equation}
\langle  \delta E_{ij}^2 \rangle = \frac{9H^4}{2\pi^2}N\,.
\label{variance}
\end{equation}
On super-Hubble scales the total two-form field is given by 
\begin{equation}
E_{ij}^{\rm classical}=E_{yz}+\delta E_{ij}\,,
\label{Eijd}
\end{equation}
with the variance (\ref{variance}) of the perturbation 
$\delta E_{ij}$.


%
\section{Statistically Anisotropic Non-gaussianity}
\label{nonsec}

In this section we estimate the statistical properties of our model, 
in particular, the scalar non-Gaussianity.
Through this section, the anisotropy is assumed to be sufficiently small. 
We derive the interacting Hamiltonian by expanding the action 
around the anisotropic background solution.
We compute correlation functions according to the in-in formalism
by neglecting the anisotropic expansion of the Universe.
The calculation is analogous to that carried out 
for the vector field in Ref.~\cite{Bartolo:2012sd}.
This prescription can be justified by more rigorous calculations
(see e.g., Ref.~\cite{Barnaby}).

\subsection{Power spectrum}

We first calculate the power spectrum of the comoving curvature 
perturbation $\zeta$ (see Refs.~\cite{zetadef} for its definition).
In our case the curvature perturbation $\zeta$ can be written as the sum of 
the ``unperturbed'' field $\zeta^{(0)}$ and the contribution 
$\delta \zeta$ coming from the two-form field.
We decompose the field $\zeta^{(0)}$ into the Fourier components
\begin{equation}
\zeta^{(0)}=\int \frac{d^3 k}{(2\pi)^{3/2}}
e^{i {\bm k}\cdot {\bm x}} \hat{\zeta}_{\bm k}^{(0)}\,,\qquad
\hat{\zeta}_k^{(0)}=\zeta_{{\bm k}}^{(0)}a_{\bm k}
+\zeta_{{\bm k}}^{(0)*}a^\dagger_{-\bm k}\,,
\label{zeta0def}
\end{equation}
where the annihilation and creation operators satisfy the 
commutation relation
\begin{equation}
[ a_{\bm k}, a^\dagger_{\bm k'} ]=
\delta^{(3)} ({\bm k}-{\bm k}')\,.
\end{equation}

At leading order in slow-roll we have the following 
solution \cite{Maldacena:2002vr}
\begin{equation}
\zeta_{{\bm k}}^{(0)}=\frac{H(1+ik\tau)}
{2\sqrt{\epsilon}M_p k^{3/2}}
e^{-ik \tau}\,.
\label{backsol}
\end{equation}
The total power spectrum ${\cal P}_{\zeta}$
is defined by the two-point correlation function of $\zeta$, as
\begin{equation}
\langle \hat{\zeta}_{{\bm k}_1} \hat{\zeta}_{{\bm k}_2} \rangle
=\frac{2\pi^2}{k_1^3} \delta^{(3)}({\bm k}_1+{\bm k}_2) 
{\cal P}_{\zeta}(k_1)\,,
\end{equation}
where $\hat{\zeta}_{{\bm k}}$ is the Fourier component of $\zeta$.
The power spectrum can be written as the sum of the two contributions 
$\zeta^{(0)}$ and $\delta \zeta$, as
\begin{equation}
{\cal P}_{\zeta}={\cal P}^{(0)}_{\zeta}
+\delta {\cal P}_{\zeta}\,.
\label{powersum}
\end{equation}
Using the solution (\ref{backsol}) long time after the Hubble radius
crossing ($\tau \to 0$), the first term 
in Eq.~(\ref{powersum}) reads
\begin{equation}
{\cal P}_{\zeta}^{(0)}=\frac{H^2}{8\pi^2 \epsilon M_p^2}\,.
\label{powerli}
\end{equation}

The next step is to derive the second contribution 
$\delta {\cal P}_{\zeta}=\delta \langle 0
|\hat{\zeta}_{{\bm k}_1} \hat{\zeta}_{{\bm k}_2} 
|0 \rangle$ from the two-form field.
The interacting Lagrangian following from 
Eq.~(\ref{actionani}) is
\begin{eqnarray}
L_{\rm int} = - \frac{a^4}{12}  \frac{\partial 
\langle f^2 \rangle }{\partial \phi} \delta \phi 
\left(H_{\mu\nu\lambda} + \delta H_{\mu\nu\lambda} \right)
\left(H^{\mu\nu\lambda} + \delta H^{\mu\nu\lambda} \right)  \ ,
\end{eqnarray}
where $\langle~~\rangle$ represents the background value.
Since the function $f$ is given by  
$f= \exp(\int d\phi /\sqrt{2\epsilon}M_p)$, 
we can deduce the following relation
\begin{eqnarray}
\frac{\partial \langle f^2 \rangle}
{\partial \phi} \delta \phi = 2 \langle f^2 \rangle  
\zeta^{(0)} \ ,
\end{eqnarray}
where we used $\delta \phi = \sqrt{2\epsilon}M_p \zeta^{(0)}$. 
Note that there is no distinction between $\epsilon$ and $\epsilon_V$
because we are considering the situation $c \sim 1$. 
Thus, we obtain
\begin{eqnarray}
  L_{\rm int} &=& \frac12 a^{-2}f^2 \left( 4 H_{0yz} \delta H_{0yz} + \delta H_{0ij} \delta H_{0ij}  \right) \zeta^{(0)}\,,\nonumber \\
  &=& \frac12 a^4 (4E_{yz} \delta E_{yz}+\delta E_{ij}\delta E_{ij}) 
  \zeta^{(0)}\,,
  \label{Lint}
\end{eqnarray}
where, in the second line, we employed the solution 
$f=a^{-1}$ and Eqs.~(\ref{Edef1})-(\ref{Edef2}).

The interacting Hamiltonian $H_{\rm int}$ is related to 
$L_{\rm int}$, as $H_{\rm int}=-\int d^3 x\,L_{\rm int}=H_1+H_2$, 
where $H_1$ and $H_2$ follow from the first and second terms
of Eq.~(\ref{Lint}).
Substituting Eqs.~(\ref{delEdef}) and (\ref{zeta0def}) 
into Eq.~(\ref{Lint}), we obtain
\begin{eqnarray}
H_1 &=& - \frac{ 2E_{yz} }{H^4 \tau^4} \int d^3 k \ 
\delta {\cal E}_{yz}(\tau , {\bm k} ) 
\hat{\zeta}_{-{\bm k}}^{(0)} (\tau)\,,  
\label{H1} \\
H_2 &=& -\frac{1}{2H^4 \tau^4 }\int \frac{d^3 k\,d^3 p}{(2\pi)^{3/2}} 
\ \delta {\cal E}_{ij}(\tau , {\bm k} ) \ \delta {\cal E}_{ij}(\tau , {\bm p} )  
\hat{\zeta}_{-{\bm k}-{\bm p}}^{(0)} (\tau ) \label{H2}\,. 
\end{eqnarray}
Using the in-in formalism \cite{Maldacena:2002vr}, the two-point 
correction following from the interacting Hamiltonian $H_1$ 
can be evaluated as
\begin{eqnarray}
\delta \langle 0|\hat{\zeta}_{{\bm k}_1} \hat{\zeta}_{{\bm k}_2} |0 \rangle 
&=& 
- \int_{\tau_{{\rm min},1}}^\tau  d\tau_1 \int_{\tau_{{\rm min},2}}^{\tau_1} d\tau_2\, \langle 0| 
\left[ \left[ \hat{\zeta}^{(0)}_{{\bm k}_1} (\tau ) \hat{\zeta}^{(0)}_{{\bm k}_2} (\tau) , H_1 (\tau_1 ) \right] , H_1 (\tau_2) \right] |0 \rangle  \nonumber\\
&=& \frac{ E_{yz}^2}{9\epsilon^2 M_p^4 H^4 } \prod_{i=1}^2 
\int_{-1/k_i}^{\tau} \frac{d\tau_i}{\tau_i^4} \left( \tau^3 - \tau_i^3 \right) 
\langle 0|\delta {\cal E}_{yz}(\tau_1,{\bm k}_1) \delta {\cal E}_{yz}
(\tau_2,{\bm k}_2) |0 \rangle  \nonumber\\
&=&  \frac{2\pi^2}{k_1^3} \delta^{(3)} ({\bm k}_1 + {\bm k}_2 )  
\frac{E_{yz}^2N_k^2 \cos^2 \theta_{{\bm k}_{1,x}}}{4\pi^2 \epsilon^2 M_p^4}\,,
\label{powerspe}
\end{eqnarray}
where $\theta_{{\bm k}_{1},x}$ is the angle between 
${\bm k}_1$ and the $x$-axis.
The two integrals in the first line of Eq.~(\ref{powerspe})
have been evaluated at the super-Hubble regime
characterized by $-k_i \tau<1$, from which 
$\tau_{{\rm min},i}=-1/k_i$ with $i=1,2$.
In the second line of Eq.~(\ref{powerspe}) the upper bound $\tau_1$ of 
the second integral has been replaced by $\tau$ by dividing the 
factor 2! because of the symmetry of the integrand. 
Note that we also used the following relation \cite{Bartolo:2012sd}
\begin{equation}
[\hat{\zeta}_{\bm k}^{(0)}(\tau), \hat{\zeta}_{\bm k'}^{(0)}(\tau')]
\simeq -\frac{iH^2 (\tau^3-\tau{'^3})}{6\epsilon M_p^2}
\delta^{(3)} ({\bm k}+{\bm k}')\,,
\end{equation}
which is valid in the super-Hubble regime.
One can show that the integral $\int_{-1/k_i}^{\tau} d\tau_i\,
(\tau^3-\tau_i^3)/\tau_i^4$ is equivalent to 
$N_{k_i} \simeq \ln (-1/k_i \tau)$ under the approximation 
$-k_i \tau \ll 1$ \cite{Bartolo:2012sd}, 
where $N_{k_i}$ is the number of e-foldings
before the end of inflation at which the modes with the wavenumber 
$k_i$ left the Hubble radius. 
Since ${\bm k}_1=-{\bm k}_2$, we used the notation $N_{k_1}=N_{k_2} 
\equiv N_k$.

Using the power spectrum (\ref{powerli})
and the slow-roll relation $3M_p^2 H^2 \simeq V$, the two-form field 
gives rise to the correction to the power spectrum
\begin{equation}
\delta {\cal P}_{\zeta}= \frac{6E_{yz}^2N_k^2}{\epsilon V}
{\cal P}_{\zeta}^{(0)} \cos^2 \theta_{{\bm k}_1,x}\,.
\label{deltaP}
\end{equation}
The interacting Hamiltonian (\ref{H2}) has a contribution 
to $\delta {\cal P}_{\zeta}$ with $E_{yz}$ replaced by the IR solution 
$\delta E_{ij}$ with the expectation value (\ref{variance}). 
Taking into account this contribution,
the total correction to the power spectrum can be 
derived by replacing $E_{yz}$ in Eq.~(\ref{deltaP}) 
for $E_{ij}^{\rm classical}$ defined in Eq.~(\ref{Eijd}).
Using the notation $E_{c} \equiv |E_{ij}^{\rm classical}|$, 
the total power spectrum of the curvature perturbation reads
\begin{equation}
{\cal P}_{\zeta} = {\cal P}_{\zeta}^{(0)} 
\left( 1  + 12 I N_k^2  \cos^2 \theta_{{\bm k}_1,x} \right)\,,
\quad {\rm where} \quad
I \equiv \frac{E_c^2}{2\epsilon V}\,.
\label{Pzeta}
\end{equation}
In contrast to the vector case where the anisotropy is
oblate \cite{Gumrukcuoglu:2010yc,Watanabe:2010fh}, 
we now have the prolate anisotropy.

The statistics of the WMAP anisotropies uses the parametrization 
${\cal P}_{\zeta} = {\cal P}_{\zeta}^{(0)}(1+g_* \cos^2 
\theta_{{\bm k},{\bm V}})$, where ${\bm V}$ is a ``privileged''
direction close to the ecliptic poles \cite{Pullen:2007tu,Carroll}.
The WMAP data provides the constraint
$g_*=0.29 \pm 0.031$ \cite{Gro}.
{}From Eq.~(\ref{Pzeta}) the anisotropy parameter is 
\begin{equation}
g_*=12 I N_k^2\,,
\end{equation}
from which we obtain
\begin{equation}
I=2.3 \times 10^{-6} 
\left(\frac{g_*}{0.1}\right) 
\left( \frac{60}{N_k} \right)^2\,.
\label{rhoE}
\end{equation}
Note that the quantity $I \epsilon=E_c^2/(2V)$ characterizes the ratio
of the energy densities of the two-form field ($E_c^2/2$) 
and inflaton ($V$), which is much smaller than 1 from 
Eq.~(\ref{rhoE}).

\subsection{Bispectrum}

The three-point correlation of $\zeta$ can be evaluated by using 
the in-in formalism along the same line of Ref.~\cite{Bartolo:2012sd}.
The tree-level contribution coming from the interacting Hamiltonian 
(\ref{H1}) is given by 
\begin{eqnarray}
   \delta \langle 0|\hat{\zeta}_{{\bm k}_1} \hat{\zeta}_{{\bm k}_2} 
   \hat{\zeta}_{{\bm k}_3} 
   |0 \rangle
   &=& i \int_{-1/k_1}^\tau  d\tau_1 \int_{-1/k_2}^{\tau_1} d\tau_2 
   \int_{-1/k_3}^{\tau_2} d\tau_3 \langle 0| 
\left[ \left[ \left[  \hat{\zeta}^{(0)}_{{\bm k}_1} \hat{\zeta}^{(0)}_{{\bm k}_2} \hat{\zeta}^{(0)}_{{\bm k}_3} (\tau) , H_2 (\tau_1 ) \right] , H_1 (\tau_2) \right] , H_1 (\tau_3) \right] |0
\rangle+{\rm 2~perm.} \nonumber\\  
   &=& \frac{E_{yz}^2}{108 \epsilon^3 M_p^6 H^6} 
   \prod_{i=1}^3 \int_{-1/k_i}^{\tau} 
   \frac{d\tau_i}{\tau_i^4} \left( \tau^3 - \tau_i^3 \right) 
   \int \frac{d^3p}{(2\pi)^{3/2}} \nonumber\\ 
     && \times \langle 0| \delta {\cal E}_{ij}(\tau_1,{\bm k}_1-{\bm p})
                          \delta {\cal E}_{ij}(\tau_1,{\bm p}) 
                          \delta {\cal E}_{yz}(\tau_2,{\bm k}_2) 
                          \delta {\cal E}_{yz}(\tau_3,{\bm k}_3) + \nonumber\\
   & &~~~~~~\delta E_{ij}(\tau_2,{\bm k}_2-{\bm p})
                          \delta {\cal E}_{ij}(\tau_2,{\bm p}) 
                          \delta {\cal E}_{yz}(\tau_3,{\bm k}_3) 
                          \delta {\cal E}_{yz}(\tau_1,{\bm k}_1) + 
   \nonumber\\
   & &~~~~~~\delta E_{ij}(\tau_3,{\bm k}_3-{\bm p})
                          \delta {\cal E}_{ij}(\tau_3,{\bm p}) 
                          \delta {\cal E}_{yz}(\tau_1,{\bm k}_1) 
                          \delta {\cal E}_{yz}(\tau_2,{\bm k}_2)
   |0 \rangle \nonumber \\                       
   &=& \frac{3 E_{yz}^2 H^2}{8 \sqrt{2} \pi^{3/2}\epsilon^3M_p^6}
   \delta^{(3)} ({\bm k}_1 + {\bm k}_2 + {\bm k}_3) N_{k_1}N_{k_2}N_{k_3}
            \left[\frac{\cos \theta_{{\bm k}_1,{\bm k}_2} \cos\theta_{{\bm k}_1,x} \cos\theta_{{\bm k}_2, x}}{k_1^3k_2^3} + \text{2 perm.}\right] \,,
            \label{bispe}
\end{eqnarray}
where ``2 perm.'' represents two terms obtained by the permutation.
In the last line of Eq.~(\ref{bispe}) we used the relation 
$\epsilon_{ij}({\bm k}_1) \epsilon_{ij}({\bm k}_2)
=2\cos \theta_{{\bm k}_1,{\bm k}_2}$.
The loop contribution following from the product of the three interacting 
Hamiltonians $H_2$ provides the bispectrum in which $E_{yz}$ of Eq.~(\ref{bispe})
is replaced by the IR solution $\delta E_{ij}$ with the variance
(\ref{variance}). 
Then the total anisotropic bispectrum ${\cal B}_{\zeta}$, defined by 
$\delta \langle 0|\hat{\zeta}_{{\bm k}_1} \hat{\zeta}_{{\bm k}_2} 
\hat{\zeta}_{{\bm k}_3} 
|0 \rangle={\cal B}_{\zeta}\delta^{(3)} ({\bm k}_1 + {\bm k}_2 + {\bm k}_3)$, 
reads
\begin{equation}
{\cal B}_{\zeta}=72\sqrt{2}\pi^{5/2}I ({\cal P}_{\zeta}^{(0)})^2 
N_{k_1}N_{k_2}N_{k_3}
\left[ \frac{\cos \theta_{{\bm k}_1,{\bm k}_2} \cos\theta_{{\bm k}_1,x} \cos\theta_{{\bm k}_2, x}}{k_1^3k_2^3} + \text{2 perm.} \right] \,,
\label{bispectrum}
\end{equation}
where $I=E_c^2/(2\epsilon V)$ is given by Eq.~(\ref{rhoE}).

We define the non-linear parameter $f_{\rm NL}$ 
according to the relation
\begin{equation}
{\cal B}_{\zeta}= \frac{3}{10} (2\pi)^{5/2} f_{\rm NL} 
({\cal P}_{\zeta})^2 \sum_{i=1}^3 k_i^3 / \prod_{i=1}^3 k_i^3\,,
\label{fnldef}
\end{equation}
by which we have
\begin{eqnarray}
f_{\rm NL}&=&60I \frac{({\cal P}_{\zeta}^{(0)})^2}
{({\cal P}_{\zeta})^2} \frac{N_{k_1} N_{k_2} N_{k_3}}{1+r_2^3+r_3^3}
[r_3^3 \cos \theta_{{\bm k}_1,{\bm k}_2} 
\cos\theta_{{\bm k}_1,x} \cos\theta_{{\bm k}_2, x}
+ \cos \theta_{{\bm k}_2,{\bm k}_3} 
\cos\theta_{{\bm k}_2,x} \cos\theta_{{\bm k}_3, x} \nonumber \\
& &~~~~~~~~~~~~~~~~~~~~~~~~~~~~~~
+r_2^3\cos \theta_{{\bm k}_3,{\bm k}_1} 
\cos\theta_{{\bm k}_3,x} \cos\theta_{{\bm k}_1, x}]\,,
\label{fnle}
\end{eqnarray}
where 
\begin{equation}
r_2 \equiv \frac{k_2}{k_1}\,,\qquad
r_3 \equiv \frac{k_3}{k_1}\,.
\end{equation}

In the following we employ the approximations 
$({\cal P}_{\zeta})^2 \simeq ({\cal P}_{\zeta}^{(0)})^2$
and $N_{k_1} \simeq N_{k_2} \simeq N_{k_3} \equiv N_{\rm CMB}$.
We also fix $r_2=1$ and define the angle $\beta=\pi-\theta_{12}$
in the range $0< \beta < \pi$ (i.e., $0 <r_3<2$).
In this case there is the following relation 
\begin{equation}
r_3^2=2(1-\cos \beta)\,.
\label{r3re}
\end{equation}
The local, equilateral, and enfolded shapes correspond to 
(i) $\beta \to 0$, $r_3 \to 0$, 
(ii) $\beta=\pi/3$, $r_3=1$, and 
(iii) $\beta \to \pi$, $r_3 \to 2$, respectively.

Let us consider the situation in which the angle between 
${\bm k}_1$ and the $x$-axis is given by $\gamma$.
On using Eq.~(\ref{rhoE}), the non-linear parameter 
(\ref{fnle}) reduces to 
\begin{equation}
f_{\rm NL} \simeq 29.8 \left( \frac{g_*}{0.1} \right)
\left( \frac{N_{\rm CMB}}{60} \right) F(r_3, \gamma)\,,
\label{fnlgene}
\end{equation}
where 
\begin{equation}
F(r_3, \gamma) \equiv \frac{1}{2+r_3^3} 
\left[ r_3^3 \cos \beta \cos \gamma 
(\cos \beta \cos \gamma+\sin \beta \sin \gamma)
+\frac12 (\cos \beta \cos \gamma +
\sin \beta \sin \gamma-\cos \gamma)^2
\right]\,.
\label{f3alpha}
\end{equation}

{}From Eq.~(\ref{r3re}) we can express $\beta$ in terms 
of $r_3$, as $\cos \beta=1-r_3^2/2$ and 
$\sin \beta=r_3 \sqrt{1-r_3^2/4}$, so that 
$f_{\rm NL}$ is a function of $r_3$ for 
a given value of $\gamma$.
The non-linear parameters for the local, equilateral, 
and enfolded shapes are given, respectively, by
\begin{eqnarray}
f_{\rm NL}^{\rm local} &=&
7.5 \left( \frac{g_*}{0.1}
\right) \left( \frac{N_{\rm CMB}}{60} \right)
\beta^2 \sin^2 \gamma \,,\label{fnl1}\\
f_{\rm NL}^{\rm equil} &=&
3.7 \left( \frac{g_*}{0.1}
\right) \left( \frac{N_{\rm CMB}}{60} \right)\,,
\label{fnl2}\\
f_{\rm NL}^{\rm enfolded} &=&
29.8 \left( \frac{g_*}{0.1}
\right) \left( \frac{N_{\rm CMB}}{60} \right)
\cos^2 \gamma \,,
\label{fnl3}
\end{eqnarray}
where, in the local case, we expanded $f_{\rm NL}^{\rm local}$ 
around $\beta=0$.
The local non-linear parameter (\ref{fnl1}) vanishes
in the squeezed limit $\beta \to 0$, which 
is a distinctive feature of our model.
The reason why this happens is that, unlike the vector models,
$f_{\rm NL}$ is proportional to the inner product of 
two vectors ${\bm k}_i$ and ${\bm k}_j$.  
In Eq.~(\ref{fnle}) the squeezed limit corresponds to the case 
in which the angles $\theta_{{\bm k}_2,{\bm k}_3}$ and 
$\theta_{{\bm k}_3,{\bm k}_1}$ approach $\pi/2$
with $r_3 \to 0$.

{}From Eq.~(\ref{fnl2}) the equilateral non-linear parameter does not 
depend on the angle $\gamma$.
For $g_*=0.3$ and $N_{\rm CMB}=60$, $f_{\rm NL}^{\rm equil}$
is as large as 10.
The enfolded non-linear parameter (\ref{fnl3}) depends on $\gamma$.
For $\cos^2 \gamma=1$, $g_*=0.1$ and $N_{\rm CMB}=60$,
$f_{\rm NL}^{\rm enfolded}$ is as large as 30.

\begin{figure}
\includegraphics[height=3.3in,width=3.5in]{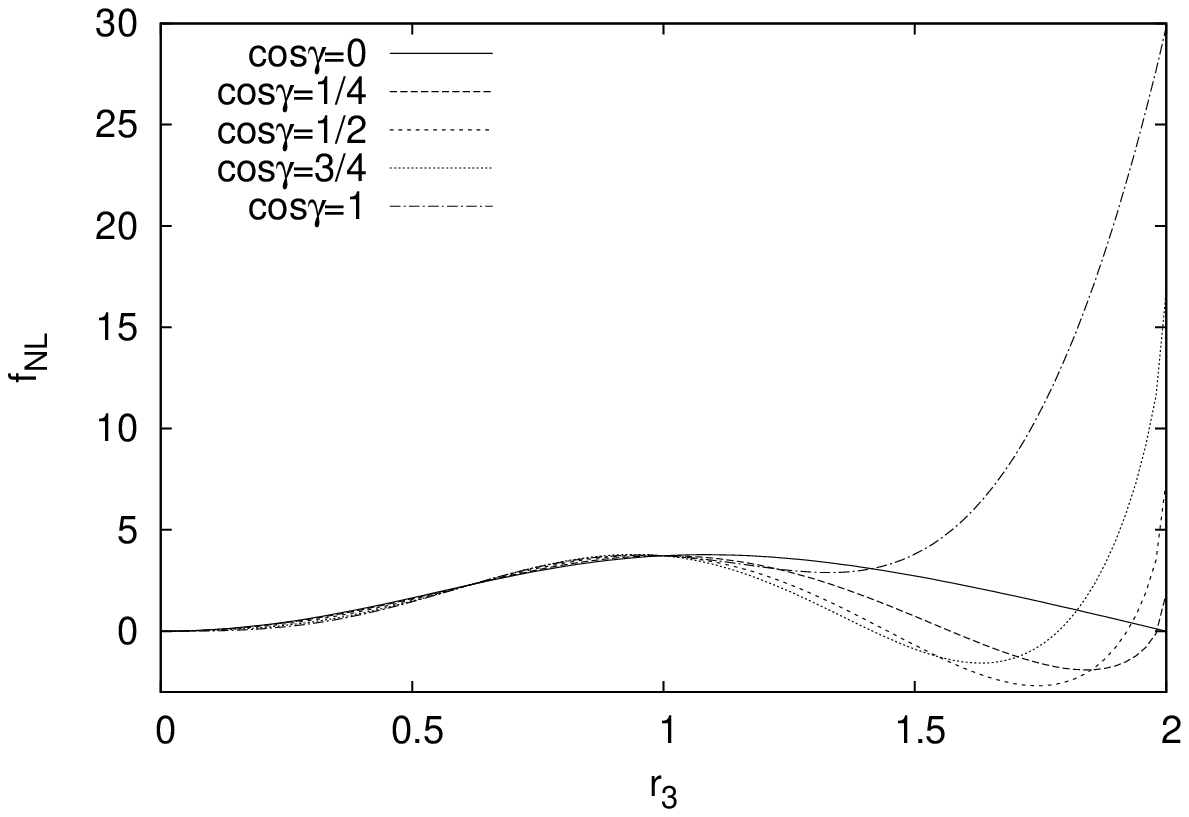}
\includegraphics[height=3.3in,width=3.5in]{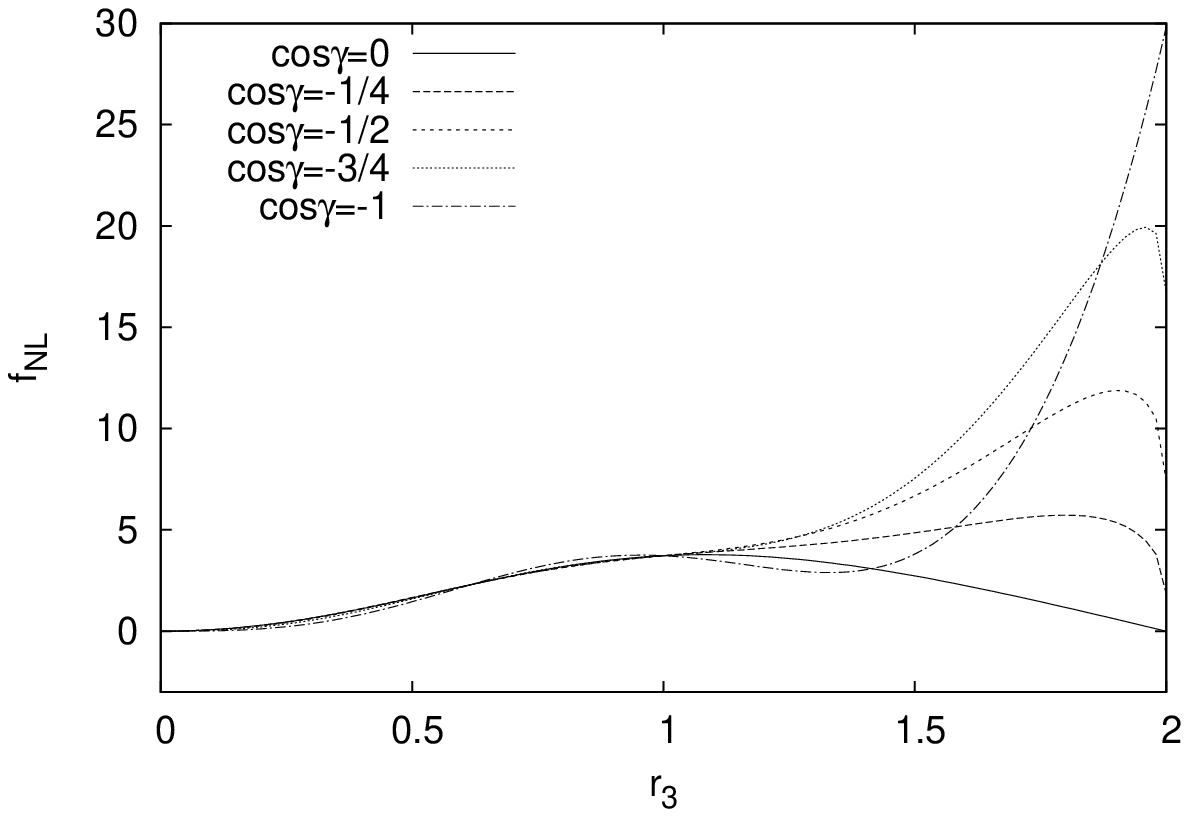}
\caption{\label{fig1}
The non-linear estimator $f_{\rm NL}$ versus $r_3=k_3/k_1$
for a number of different values of $\cos \gamma$
with $g_*=0.1$ and $N_{\rm CMB}=60$.
The left and right panels show the plots for 
the angles $0 \le \gamma \le \pi/2$ and 
$\pi/2 \le \gamma \le \pi$, respectively. 
The local, equilateral, and enfolded limits correspond 
to $r_3=0$, $r_3=1$, and $r_3=2$, respectively.
For $\gamma$ close to $\pi/2$, the equilateral 
non-linear parameter is largest.
For $\gamma$ close to 0 or $\pi$, $f_{\rm NL}$
has a maximum at $r_3=2$.}
\end{figure}

In Fig.~\ref{fig1} we plot the non-linear parameter 
(\ref{fnlgene}) versus $r_3$ ($0<r_3<2$)
for $g_*=0.1$ and $N_{\rm CMB}=60$.
The left panel and right panel correspond to 
positive and negative values of $\cos \gamma$, respectively.
In the local limit ($r_3 \to 0$), the estimator $f_{\rm NL}$ 
vanishes for any value of $\gamma$.
For the angle $\gamma$ close to $\pi/2$, $f_{\rm NL}$
has a maximum at the equilateral configuration ($r_3=1$).
With the increase of $|\cos \gamma|$, however, the enfolded estimator
gets larger. In particular, for $\gamma$ close to $0$ or $\pi$, 
$f_{\rm NL}$ has a maximum at $r_3=2$.

%
\subsection{Trispectrum}

The four-point correlation function of $\zeta$ corresponding 
to the tree-level contribution is given by 
\begin{eqnarray}
   \delta \langle 0|\hat{\zeta}_{{\bm k}_1} \hat{\zeta}_{{\bm k}_2} 
   \hat{\zeta}_{{\bm k}_3} \hat{\zeta}_{{\bm k}_4}
   |0 \rangle
   &=& \int_{-1/k_1}^\tau  d\tau_1 \int_{-1/k_2}^{\tau_1} d\tau_2 
   \int_{-1/k_3}^{\tau_2} d\tau_3 \int_{-1/k_4}^{\tau_3} d\tau_4 \nonumber \\
   &&\times
   \langle 0| 
\left[ \left[ \left[ \left[  \hat{\zeta}^{(0)}_{{\bm k}_1} 
\hat{\zeta}^{(0)}_{{\bm k}_2} 
\hat{\zeta}^{(0)}_{{\bm k}_3} \hat{\zeta}^{(0)}_{{\bm k}_4} (\tau) , H_2 (\tau_1 ) \right] , H_2 (\tau_2) \right] , H_1 (\tau_3) \right]
 , H_1 (\tau_4) \right] |0
\rangle+{\rm 5~perm.} \nonumber\\  
   &=& \frac{1}{2 \cdot 6^4}
   \frac{E_{yz}^2}{H^8 \epsilon^4 M_p^8}
   \prod_{i=1}^4 \int_{-1/k_i}^{\tau}
    \frac{d\tau_i}{\tau_i^4} \left( \tau^3 - \tau_i^3 \right) 
   \int \frac{d^3pd^3 p'}{(2\pi)^{3}} \nonumber\\ 
     && \times \langle 0| \delta {\cal E}_{ij}(\tau_1,{\bm k}_1-{\bm p})
                          \delta {\cal E}_{ij}(\tau_1,{\bm p}) 
                          \delta {\cal E}_{ij}(\tau_2,{\bm k}_2-{\bm p}')
                          \delta {\cal E}_{ij}(\tau_2,{\bm p}')
                          \delta {\cal E}_{yz}(\tau_3,{\bm k}_3) 
                          \delta {\cal E}_{yz}(\tau_4,{\bm k}_4) \nonumber \\
                       & &~~~~~+  11~{\rm perms.}|0 \rangle \nonumber \\                       
   &=& -\frac{9}{8 \cdot (2\pi)^3} \frac{E_{yz}^2 H^4}{\epsilon^4 M_p^8}
   \delta^{(3)} ({\bm k}_1 + {\bm k}_2 + {\bm k}_3+ {\bm k}_4) 
            N_{k_1}N_{k_2}N_{k_3}N_{k_4} \nonumber \\
   && \times  \left[ \frac{1}{k_1^3 k_2^3 k_{13}^3}
            \cos \theta_{{\bm k}_1,{\bm k}_{13}} 
            \cos \theta_{{\bm k}_2,{\bm k}_{13}}
            \cos \theta_{{\bm k}_1,x}
             \cos \theta_{{\bm k}_2,x}
            + \text{11 perm.}\right] \,,
            \label{trispe}
\end{eqnarray}
where ${\bm k}_{ij}={\bm k}_i+{\bm k}_j$.
The loop contribution, which follows from the product of the 
four interacting Hamiltonians $H_2$, gives rise to the four-point 
correlation function where $E_{yz}$ in Eq.~(\ref{trispe}) is 
replaced by the IR solution $\delta E_{ij}$ with 
the variance (\ref{variance}).
Defining the total anisotropic trispectrum ${\cal T}_{\zeta}$, 
as $\delta \langle 0|\hat{\zeta}_{{\bm k}_1} \hat{\zeta}_{{\bm k}_2} 
\hat{\zeta}_{{\bm k}_3}
\hat{\zeta}_{{\bm k}_4} |0 \rangle
={\cal T}_{\zeta}\,\delta^{(3)} ({\bm k}_1 + {\bm k}_2 + {\bm k}_3 + {\bm k}_4)$, 
we obtain
\begin{equation}
{\cal T}_{\zeta}=-432 \pi^3 I (P_{\zeta}^{(0)})^3
N_{k_1} N_{k_2} N_{k_3} N_{k_4}
\left[ \frac{1}{k_1^3 k_2^3 k_{13}^3}
            \cos \theta_{{\bm k}_1,{\bm k}_{13}} 
            \cos \theta_{{\bm k}_2,{\bm k}_{13}}
            \cos \theta_{{\bm k}_1,x}
             \cos \theta_{{\bm k}_2,x}
            + \text{11 perm.}\right]\,.
\label{Tzetaf}
\end{equation}
We introduce the non-linear estimator $\tau_{\rm NL}$
according to the relation
\begin{equation}
{\cal T}_{\zeta}=(2\pi)^3 ({\cal P}_{\zeta})^3
\frac{\tau_{\rm NL}}{8} \left( \frac{1}
{k_1^3 k_2^3 k_{13}^3}+\text{11 perm.} \right)\,.
\label{taudef}
\end{equation}

In the squeezed limit characterized by $k_{12} \to 0$, 
the non-linear estimator reduces to
\begin{eqnarray}
\tau_{\rm NL}^{\rm local} 
&=& -108 I N_{k_1}^2 N_{k_3}^2 
[  \cos \theta_{{\bm k}_1,{\bm k}_{12}} 
   \cos \theta_{{\bm k}_3,{\bm k}_{12}}
   \cos \theta_{{\bm k}_1,x}
   \cos \theta_{{\bm k}_3,x}+
   \cos \theta_{{\bm k}_1,{\bm k}_{12}} 
   \cos \theta_{{\bm k}_4,{\bm k}_{12}}
   \cos \theta_{{\bm k}_1,x}
   \cos \theta_{{\bm k}_4,x} \nonumber \\
   &&~~~~~~~~~~~~~~~~~+
   \cos \theta_{{\bm k}_2,{\bm k}_{12}} 
   \cos \theta_{{\bm k}_3,{\bm k}_{12}}
   \cos \theta_{{\bm k}_2,x}
   \cos \theta_{{\bm k}_3,x}+
   \cos \theta_{{\bm k}_2,{\bm k}_{12}} 
   \cos \theta_{{\bm k}_4,{\bm k}_{12}}
   \cos \theta_{{\bm k}_2,x}
   \cos \theta_{{\bm k}_4,x} ]\,,
\end{eqnarray}
where we used the approximation 
$({\cal P}_{\zeta})^3 \simeq ({\cal P}_{\zeta}^{(0)})^3$.
Since the angles between the vectors ${\bm k}_i$ ($i=1,2,3,4$)
and ${\bm k}_{12}$ approach $\pi/2$ for $k_{12} \to 0$, 
the estimator $\tau_{\rm NL}^{\rm local}$ vanishes
in this limit.

We also consider the regular tetrahedron limit, i.e., 
$k_1=k_2=k_3=k_4=k_{12}=k_{14} \equiv k$ (see e.g., 
figure 2 of Ref.~\cite{CHHSW} for illustration).
For this configuration, the angle between ${\bm k}_{13}$
and ${\bm k}_{1}$ is $\pi/4$ with $k_{13}=\sqrt{2}k$.
We also focus on the case in which the direction of ${\bm k}_1$ 
is the same as that of the $x$-axis. 
Then the trispectrum (\ref{Tzetaf}) reads 
\begin{equation}
{\cal T}_{\zeta}^{\rm equil} \simeq -54(\sqrt{2}+1) \pi^3 I 
({\cal P}_{\zeta}^{(0)})^3 N_{k_1}^4
\frac{1}{k^9}\,,
\end{equation}
by which the non-linear estimator can be derived as
\begin{equation}
\tau_{\rm NL}^{\rm equil} \simeq -4.1 \times 10^{2}
\left( \frac{g_*}{0.1} \right) \left( \frac{N_{k_1}}
{60} \right)^2\,.
\end{equation}
Unlike the local shape, $|\tau_{\rm NL}^{\rm equil}|$
can be of the order of $10^2$-$10^3$.


%
\section{Conclusions}
\label{consec}

For the models in which the inflaton field $\phi$ couples to an 
anti-symmetric tensor $B_{\mu \nu}$, we showed that 
anisotropic inflation occurs for the coupling $f(\phi)$ 
given by Eq.~(\ref{eq:function}).
In this case there is an attractor solution along which the 
ratio of the anisotropic shear $\Sigma$ to the Hubble parameter 
$H$ is proportional to the slow-roll parameter $\epsilon$.
Even for the super-critical case in which the coupling $f(\phi)$
is generalized to Eq.~(\ref{formula:f}) with $c>1$, 
there is the regime of anisotropic inflation where $\Sigma/H$
is nearly constant with $f$ proportional to $a^{-1}$.
The anisotropy induced by the two-form field corresponds to 
the prolate type (the expansion of the 
Universe slows down in the ($y,z$) plane), 
in contrast to the oblate type stemming 
from the vector field.

The presence of the two-form field coupled to inflaton gives
rise to modifications to statistical quantities 
observed in CMB temperature fluctuations.
{}From the action (\ref{actionani}) we derived the interacting Hamiltonians
(\ref{H1}) and (\ref{H2}) between curvature perturbations and 
the two-form field.
We evaluated the $n$-point correlation functions ($n=2,3,4$) 
of curvature perturbations by using the in-in formalism of 
quantum field theory.
The 2-point correlation function, i.e., 
the power spectrum, is given by Eq.~(\ref{Pzeta}), where 
$g_*=12I N_k^2$ parametrizes the strength of anisotropy.
Even if the energy density of the two-form field is very much 
smaller than that of inflaton, the parameter $g_*$ can be 
of the order of 0.1, as suggested by the WMAP data \cite{Gro}.

In Eqs.~(\ref{bispectrum}) and (\ref{fnle}) we derived 
the three-point correlation function 
${\cal B}_{\zeta}$ (bispectrum) and the non-linear estimator $f_{\rm NL}$, 
which exhibit a number of interesting properties.
By considering the triangle of three momenta (${\bm k}_1+{\bm k}_2+{\bm k}_3=0$) 
with $k_1=k_2$, we showed that $f_{\rm NL}$ can be expressed by 
functions of $r_3=k_3/k_1$ and the angle $\gamma$ between ${\bm k}_1$
and the $x$-axis. 
In the local, equilateral, and enfolded limits, the non-linear 
estimators are simplify given by Eqs.~(\ref{fnl1}), (\ref{fnl2}), 
and (\ref{fnl3}), respectively.
We found that $f_{\rm NL}^{\rm local}$ vanishes in the squeezed
limit ($r_3 \to 0$), whereas $f_{\rm NL}^{\rm equil}$ and 
$f_{\rm NL}^{\rm enfolded}$ can be of the order of 10 
(see Fig.~\ref{fig1}).
These results are consistent with the recent constraints by Planck, 
i.e., $f_{\rm NL}^{\rm local}=2.7 \pm 5.8$ and 
$f_{\rm NL}^{\rm equil}=-42 \pm 75$ (68~\%\,CL) \cite{Planckfirst}. 
 
The four-point correlation function ${\cal T}_{\zeta}$ 
(trispectrum) has been also computed in Eq.~(\ref{Tzetaf}).
Defining the non-linear estimator $\tau_{\rm NL}$ as 
Eq.~(\ref{taudef}), we found that $\tau_{\rm NL}$ vanishes
in the squeezed limit ($k_{12} \to 0$).
However, for other shapes such as the regular tetrahedron, 
$|\tau_{\rm NL}|$ can be of the order of $10^2$-$10^3$.
This is an interesting property by which our scenario can be
distinguished from the vector case as well as other models
with large non-Gaussianities.

It will be of interest to understand the physics of 
a dipole-type anisotropy suggested by the Planck data~\cite{Planckfirst}.
Although there is a phenomenological description of 
this type of anisotropy~\cite{Gordon:2005ai},
no physically well-motivated models are present to our best knowledge. 
Recent attempt to explain the dipole-type anisotropy
by a contrived geometrical set up is intriguing~\cite{Chang:2013vla},
but it still lacks a consistent dynamical picture.
Since our framework is natural and consistent, it would be great 
if our mechanism is generalized to explain the origin of 
the dipole-type anisotropy observed by Planck.

\acknowledgements

This work is supported by the Grant-in-Aid for Scientific Research 
Fund of the Ministry of Education, Science and Culture of Japan 
(Nos.~23$\cdot$6781, 22540274, and 24540286), the Grant-in-Aid 
for Scientific Research on Innovative Area (No.~21111006).


\begin{thebibliography}{99}

\bibitem{infpapers} 
A.~A.~Starobinsky,
Phys.\ Lett.\ B {\bf 91}, 99 (1980);\\
D.~Kazanas,
Astrophys.\ J.\  {\bf 241} L59 (1980);\\
K.~Sato, Mon.\ Not.\ R.\ Astron.\ Soc. {\bf 195}, 
467 (1981);\\
A.~H.~Guth,
Phys.\ Rev.\ D {\bf 23}, 347 (1981).

\bibitem{infper}
V.~F.~Mukhanov and G.~V.~Chibisov,
JETP Lett.\  {\bf 33}, 532 (1981);\\
A.~H.~Guth and S.~Y.~Pi,
Phys.\ Rev.\ Lett.\  {\bf 49}, 1110 (1982);\\
S.~W.~Hawking,
Phys.\ Lett.\ B {\bf 115}, 295 (1982);\\
A.~A.~Starobinsky,
Phys.\ Lett.\ B {\bf 117}, 175 (1982).

\bibitem{review}
J.~E.~Lidsey {\it et al.},
Rev.\ Mod.\ Phys.\  {\bf 69}, 373 (1997)
[astro-ph/9508078];\\
D.~H.~Lyth and A.~Riotto,
Phys.\ Rept.\  {\bf 314}, 1 (1999)
[hep-ph/9807278].

\bibitem{WMAP1}
D.~N.~Spergel {\it et al.}  [WMAP Collaboration],
Astrophys.\ J.\ Suppl.\  {\bf 148}, 175 (2003)
[astro-ph/0302209].

\bibitem{WMAP9} 
C.~L.~Bennett {\it et al.},
arXiv:1212.5225 [astro-ph.CO];\\
G.~Hinshaw {\it et al.},
arXiv:1212.5226 [astro-ph.CO].

\bibitem{Planckfirst} 
P.~A.~R.~Ade {\it et al.}  [ Planck Collaboration],
arXiv:1303.5076 [astro-ph.CO];
arXiv:1303.5082 [astro-ph.CO];
arXiv:1303.5083 [astro-ph.CO];
arXiv:1303.5084 [astro-ph.CO].

\bibitem{Yokoyama}
S.~Yokoyama and J.~Soda,
JCAP {\bf 0808}, 005 (2008)
[arXiv:0805.4265 [astro-ph]].

\bibitem{Dimopoulos}
K.~Dimopoulos, M.~Karciauskas, D.~H.~Lyth and Y.~Rodriguez,
JCAP {\bf 0905}, 013 (2009)
[arXiv:0809.1055 [astro-ph]];\\
M.~Karciauskas, K.~Dimopoulos and D.~H.~Lyth,
Phys.\ Rev.\  D {\bf 80} (2009) 023509
[arXiv:0812.0264 [astro-ph]];\\
C.~A.~Valenzuela-Toledo, Y.~Rodriguez and D.~H.~Lyth,
Phys.\ Rev.\  D {\bf 80}, 103519 (2009)
[arXiv:0909.4064 [astro-ph.CO]];\\
C.~A.~Valenzuela-Toledo and Y.~Rodriguez,
Phys.\ Lett.\  B {\bf 685}, 120 (2010)
[arXiv:0910.4208 [astro-ph.CO]];\\
K.~Dimopoulos,
Int.\ J.\ Mod.\ Phys.\ D {\bf 21}, 1250023 (2012)
[Erratum-ibid.\ D {\bf 21}, 1292003 (2012)]
[arXiv:1107.2779 [hep-ph]];\\
N.~Bartolo, E.~Dimastrogiovanni, S.~Matarrese and A.~Riotto,
JCAP {\bf 0910}, 015 (2009)
[arXiv:0906.4944 [astro-ph.CO]];\\
N.~Bartolo, E.~Dimastrogiovanni, S.~Matarrese and A.~Riotto,
JCAP {\bf 0911}, 028 (2009)
[arXiv:0909.5621 [astro-ph.CO]];\\
E.~Dimastrogiovanni, N.~Bartolo, S.~Matarrese and A.~Riotto,
Adv.\ Astron.\  {\bf 2010}, 752670 (2010)
[arXiv:1001.4049 [astro-ph.CO]];\\
R.~Emami and H.~Firouzjahi,
JCAP {\bf 1201}, 022 (2012)
[arXiv:1111.1919 [astro-ph.CO]];\\
M.~Karciauskas,
JCAP {\bf 1201}, 014 (2012)
[arXiv:1104.3629 [astro-ph.CO]];\\
M.~Shiraishi and S.~Yokoyama,
Prog.\ Theor.\ Phys.\  {\bf 126}, 923 (2011)
[arXiv:1107.0682 [astro-ph.CO]].

\bibitem{Watanabe}
M.~a.~Watanabe, S.~Kanno and J.~Soda,
Phys.\ Rev.\ Lett.\  {\bf 102}, 191302 (2009)
[arXiv:0902.2833 [hep-th]].

\bibitem{Gumrukcuoglu:2010yc} 
A.~E.~Gumrukcuoglu, B.~Himmetoglu and M.~Peloso,
Phys.\ Rev.\ D {\bf 81}, 063528 (2010)
[arXiv:1001.4088 [astro-ph.CO]];\\
T.~R.~Dulaney, M.~I.~Gresham and ,
Phys.\ Rev.\ D {\bf 81}, 103532 (2010)
[arXiv:1001.2301 [astro-ph.CO]].

\bibitem{Watanabe:2010fh}
M.~a.~Watanabe, S.~Kanno and J.~Soda,
Prog.\ Theor.\ Phys.\  {\bf 123}, 1041 (2010)
[arXiv:1003.0056 [astro-ph.CO]].  

\bibitem{Emami:2013bk} 
R.~Emami and H.~Firouzjahi,
arXiv:1301.1219 [hep-th].

\bibitem{Soda:2012zm} 
J.~Soda,
Class.\ Quant.\ Grav.\  {\bf 29}, 083001 (2012)
[arXiv:1201.6434 [hep-th]].

\bibitem{Maleknejad} 
A.~Maleknejad, M.~M.~Sheikh-Jabbari and J.~Soda,
arXiv:1212.2921 [hep-th].

\bibitem{Watanabe2}
M.~a.~Watanabe, S.~Kanno and J.~Soda,
Mon.\ Not.\ Roy.\ Astron.\ Soc.\  {\bf 412}, L83 (2011)
[arXiv:1011.3604 [astro-ph.CO]].
 
\bibitem{BarnabyPRL} 
N.~Barnaby and M.~Peloso, 
Phys.\ Rev.\ Lett.\  {\bf 106}, 181301 (2011)
[arXiv:1011.1500 [hep-ph]];\\
N.~Barnaby, R.~Namba and M.~Peloso,
JCAP {\bf 1104}, 009 (2011)
[arXiv:1102.4333 [astro-ph.CO]];\\
N.~Barnaby, E.~Pajer and M.~Peloso, 
Phys.\ Rev.\ D {\bf 85}, 023525 (2012)
[arXiv:1110.3327 [astro-ph.CO]].

\bibitem{Barnaby}
N.~Barnaby, R.~Namba and M.~Peloso,
Phys.\ Rev.\ D {\bf 85}, 123523 (2012)
[arXiv:1202.1469 [astro-ph.CO]].

\bibitem{Bartolo:2012sd}
N.~Bartolo, S.~Matarrese, M.~Peloso and A.~Ricciardone,
Phys.\ Rev.\ D {\bf 87} (2013) 023504
[arXiv:1210.3257 [astro-ph.CO]].

\bibitem{Shiraishi} 
M.~Shiraishi, E.~Komatsu, M.~Peloso and N.~Barnaby,
arXiv:1302.3056 [astro-ph.CO].

\bibitem{Abol} 
A.~A.~Abolhasani, R.~Emami, J.~T.~Firouzjaee and H.~Firouzjahi,
arXiv:1302.6986 [astro-ph.CO].

\bibitem{Lyth13} 
D.~H.~Lyth and M.~Karciauskas,
arXiv:1302.7304 [astro-ph.CO].

\bibitem{Baghram:2013lxa} 
  S.~Baghram, M.~H.~Namjoo and H.~Firouzjahi,
  arXiv:1303.4368 [astro-ph.CO].

\bibitem{Kanno:2010nr}
S.~Kanno, J.~Soda and M.~a.~Watanabe,
JCAP {\bf 1012}, 024 (2010)
[arXiv:1010.5307 [hep-th]];\\
K.~Murata and J.~Soda,
JCAP\ {\bf 1106}, 037  (2011)
[arXiv:1103.6164 [hep-th]];\\
T.~Q.~Do, W.~F.~Kao and I.~-C.~Lin,
Phys.\ Rev.\ D\ {\bf 83}, 123002  (2011);\\
T.~Q.~Do and W.~F.~Kao,
Phys.\ Rev.\  D {\bf 84}, 123009 (2011);\\
R.~Emami, H.~Firouzjahi, S.~M.~Sadegh Movahed and M.~Zarei,
JCAP {\bf 1102}, 005 (2011)
[arXiv:1010.5495 [astro-ph.CO]];\\
P.~V.~Moniz and J.~Ward, 
Class.\ Quant.\ Grav.\  {\bf 27}, 235009 (2010)
[arXiv:1007.3299 [gr-qc]];\\
S.~Bhowmick and S.~Mukherji,
Mod.\ Phys.\ Lett.\ A {\bf 27}, 1250009 (2012)
[arXiv:1105.4455 [hep-th]];\\
S.~Hervik, D.~F.~Mota and M.~Thorsrud,
JHEP {\bf 1111}, 146 (2011)
[arXiv:1109.3456 [gr-qc]];\\
K.~Yamamoto, M.~a.~Watanabe and J.~Soda,
Class.\ Quant.\ Grav.\  {\bf 29}, 145008 (2012)
[arXiv:1201.5309 [hep-th]].

\bibitem{LCW} 
J.~E.~Lidsey, D.~Wands and E.~J.~Copeland,
Phys.\ Rept.\  {\bf 337}, 343 (2000)
[hep-th/9909061].

\bibitem{Kanno:2009ei}
S.~Kanno, J.~Soda and M.~a.~Watanabe,
JCAP {\bf 0912}, 009 (2009)
[arXiv:0908.3509 [astro-ph.CO]];\\
J.~M.~Wagstaff and K.~Dimopoulos,
Phys.\ Rev.\  D {\bf 83}, 023523 (2011)
[arXiv:1011.2517 [hep-ph]].

\bibitem{zetadef} 
V.~F.~Mukhanov, H.~A.~Feldman and R.~H.~Brandenberger,
Phys.\ Rept.\  {\bf 215}, 203 (1992);\\
B.~A.~Bassett, S.~Tsujikawa and D.~Wands,
Rev.\ Mod.\ Phys.\  {\bf 78}, 537 (2006)
[astro-ph/0507632].

\bibitem{Maldacena:2002vr}
J.~M.~Maldacena,
JHEP {\bf 0305}, 013 (2003).

\bibitem{Pullen:2007tu} 
A.~R.~Pullen and M.~Kamionkowski,
Phys.\ Rev.\ D {\bf 76}, 103529 (2007)
[arXiv:0709.1144 [astro-ph]].

\bibitem{Carroll} 
L.~Ackerman, S.~M.~Carroll and M.~B.~Wise, 
Phys.\ Rev.\ D {\bf 75}, 083502 (2007)
[Erratum-ibid.\ D {\bf 80}, 069901 (2009)]
[astro-ph/0701357].

\bibitem{Gro} 
N.~E.~Groeneboom, L.~Ackerman, I.~K.~Wehus and H.~K.~Eriksen,
Astrophys.\ J.\  {\bf 722}, 452 (2010)
[arXiv:0911.0150 [astro-ph.CO]].

\bibitem{CHHSW} 
X.~Chen, B.~Hu, M.~-x.~Huang, G.~Shiu and Y.~Wang,
JCAP {\bf 0908}, 008 (2009)
[arXiv:0905.3494 [astro-ph.CO]].

\bibitem{Gordon:2005ai} 
C.~Gordon, W.~Hu, D.~Huterer and T.~M.~Crawford,
Phys.\ Rev.\ D {\bf 72}, 103002 (2005)
[astro-ph/0509301].

\bibitem{Chang:2013vla} 
Z.~Chang and S.~Wang,
arXiv:1303.6058 [astro-ph.CO].

\end{thebibliography}
\end{document}